\documentclass[pdflatex,sn-mathphys]{sn-jnl}% Math and Physical Sciences Reference Style
\jyear{2022}%

\theoremstyle{thmstyleone}%
\theoremstyle{thmstyletwo}%

\theoremstyle{thmstylethree}%
\usepackage{comment}

\usepackage{xcolor}

\newcommand{\new}[1]    {{#1}}

\begin{document}

\title[Continuous Design Control for ML in Certified Medical Systems]{Continuous Design Control for Machine Learning in Certified Medical Systems}


\author*[1,5]{\fnm{Vlad} \sur{Stirbu}}\email{vlad.stirbu@compliancepal.eu}

\author[2,3]{\fnm{Tuomas} \sur{Granlund}}\email{tuomas.granlund@solita.fi}
\equalcont{These authors contributed equally to this work.}

\author[4,5]{\fnm{Tommi} \sur{Mikkonen}}\email{tommi.j.mikkonen@jyu.fi, tommi.mikkonen@helsinki.fi}
\equalcont{These authors contributed equally to this work.}

\affil*[1]{%\orgdiv{Department}, 
\orgname{CompliancePal}, \orgaddress{%\street{Street}, 
\city{Tampere}, %\postcode{100190}, \state{State}, 
\country{Finland}}}

\affil[2]{
%\orgdiv{Department}, 
\orgname{Solita}, \orgaddress{%\street{Street}, 
\city{Tampere}, %\postcode{10587}, \state{State}, 
\country{Finland}}}

\affil[3]{%\orgdiv{Department}, 
\orgname{Tampere University}, \orgaddress{ %\street{Street}, 
\city{Tampere}, %\postcode{610101}, \state{State}, 
\country{Finland}}}

\affil[4]{%\orgdiv{Department}, 
\orgname{University of Jyväskylä}, \orgaddress{ %\street{Street}, 
\city{Jyväskylä}, %\postcode{610101}, \state{State}, 
\country{Finland}}}

\affil[5]{%\orgdiv{Department}, 
\orgname{University of Helsinki}, \orgaddress{ %\street{Street}, 
\city{Helsinki}, %\postcode{610101}, \state{State}, 
\country{Finland}}}


\abstract{Continuous software engineering has become commonplace in numerous fields. However, in regulating intensive sectors, where additional concerns needs to be taken into account, it is often considered difficult to apply continuous development approaches, such as devops. In this paper, we present an approach for using pull requests as design controls, and apply this approach to machine learning in certified medical systems leveraging model cards, a novel technique developed to add explainability to machine learning systems, as a regulatory audit trail. The approach is demonstrated with an industrial system that we have used previously to show how medical systems can be developed in a continuous fashion.}

\keywords{Machine learning, ML, MLOps, CD4ML, design control, medical software, regulated software, continuous engineering}



\maketitle

\section{Introduction}

During the latest decade, the Web has silently become the dominant platform for software applications. Effectively, this process has made releasing software so simple and cheap that to a degree, development and deployment activities are entangled. New parts of software are experimentally deployed, and feedback from released software is used to assist in development. As pointed out in \cite{taivalsaari2008web}, this leads to a new type of development approach, advancing in evolutionary fashion, where software is always on, and updates are tiny changes in the code. Concrete models for such continuous software engineering \cite{fitzgerald2017continuous} include continuous delivery \cite{humble2010continuous} and DevOps \cite{ebert2016devops}.

However, not all software lives on the Web, where applications can constantly evolve behind the curtains. Instead, numerous applications are used to power factories, control electronics, provide guidance for drones, and so on. For these, it is common that additional concerns are added in the development process. These can be added to the continuous development process as add-ons or amalgamations, sometimes also reflected in their respective names, such as DevSecOps \cite{myrbakken2017devsecops} for DevOps used to develop secure systems, RegOps \cite{drvar2020future} for digitalizing the regulatory value chain, or MLOps \cite{treveil2020introducing} for continuous delivery of Machine Learning (ML) features. 

\begin{figure}[htb]
    \centering
    \includegraphics[width=0.9\textwidth]{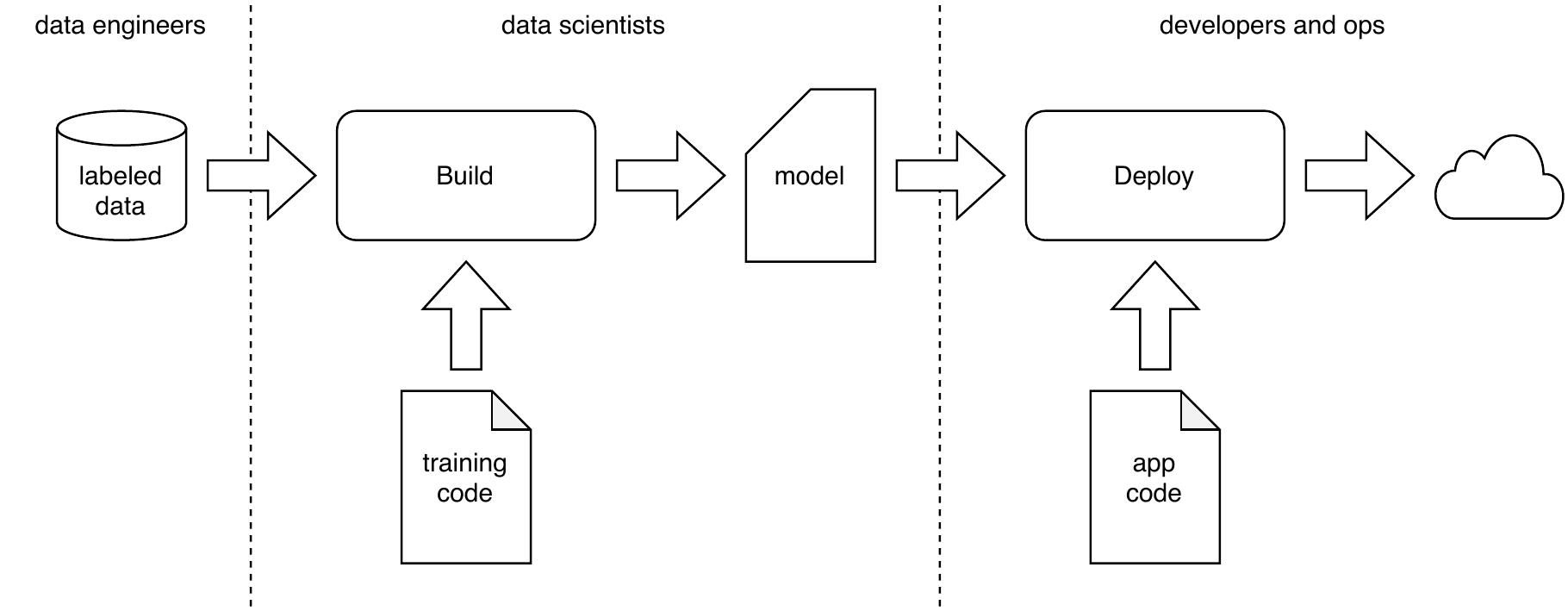}
    \caption{Simplified MLOps pipeline. Figure adapted from \cite{oravizio}.}
    \label{fig:ml-silo}
\end{figure}

Unfortunately, these approaches focus on one particular aspect that is added to the continuous software engineering pipeline, and do not consider how to integrate them to a bigger whole. Hence, their interoperability remains weak.
Consequently, relating regulatory compliance and ML, for instance, requires additional considerations which are not a part of any off-the-shelf approach. As an example, consider the MLOps pipeline visualized in Figure \ref{fig:ml-silo}, consisting of data operations, executed by data engineers, data analysis and ML operations, run by data scientists, and developers who implement and deploy the final application. In contrast, RegOps focuses only on software development -- the final part, performed by software developers -- but overlooks the rest \cite{toivakka,stirbu2018towards}. Hence, while MLOps helps in forming a pipeline for the whole development effort, RegOps only supports the final parts with regulatory considerations and design controls.

In this paper, we introduce continuous design controls for ML in certified medical systems, covering the MLOps pipeline. The proposed approach starts with continuous software engineering practices, which is then expanded with ML and data processing facilities. Then, we introduce the necessary regulatory processes, which cover both software and ML parts of the development. To simplify presentation, details of data operations, which in any case are often specific to certain organisation \cite{aho2020demystifying}, are largely overlooked, although practical techniques that bind them to continuous software engineering practices are included in the paper. The resulting model is then demonstrated with an industrial case study that we have used in our previous paper \cite{oravizio}, with an extended discussion regarding the proposed improvements.

The rest of this paper is structured as follows. In Section 2, we introduce the necessary background of the paper. In Section 3, we propose a solution for continuous design controls. In Section 4, we demonstrate the solution with an industry example. In Section 5, we discuss the implications of the proposed solution. In Section 6, we draw some final conclusions.

\section{Background}

Because of multi-faceted nature of this work, it  combines several different research fields, including continuous software development, ML, and the landscape of medical regulations. In the following, we present recent advances in these fields, so that we can introduce the proposed practical pipeline for regulated MLOps. 

\subsection{Continuous software engineering practices}

The core of continuous software engineering practices is two-fold. On one hand, it consists of a mindset where the developers take responsibility for the whole software as a whole, and, while a single developer works on a particular feature, the bigger whole is constructed in terms of developers' collaborative effort. On the other hand, it includes a toolset that allows deploying new features to use as soon as they are available \cite{fitzgerald2017continuous}. The goal is to produce continuous flow of value adding software artifacts from the  development to the actual production use, with quality assurance also happening continuously as a part of the flow. 

A particular flavor of continuous software engineering is called DevOps \cite{debois2011devops,rajkumar2016devops}. It can be described as a set of practices whose goal is to shorten the commit feedback cycle without compromising quality \cite{bass2015devops}, and to expand the development team with the operators. In addition, the continuous software engineering toolset is expanded with monitoring capabilities, ensuring that each stakeholder gets a timely access to what they need. 

Finally, it is important to notice that the automated pipeline is not about software going into production without any operator supervision, but rather the pipeline provides a feedback loop to each of the stakeholders from all stages of the delivery process. Moreover, as the software progresses through the pipeline, different stages can be triggered for example by operations and test teams by the click of a button.

\subsection{ML lifecycle challenges}

The typical process of developing a ML application starts with transforming input data from a variety of sources into a collection of labeled data, a procedure performed by data engineers. This collection is then used by data scientists to perform a series of experiments that allows then to build a set of candidate models. The model that fulfils best the desired criteria captured in the functional and non-functional requirements is selected for deployment. Finally, the selected model is incorporated into the software system and deployed to the production environment, a procedure that is performed by software and operations engineers.

Using continuous software engineering practices throughout ML application lifecycle is not straightforward. In Fig. \ref{fig:ml-silo}, we can see that the development is performed across three competence clusters: data engineers, data scientist and software developers. The level of expertise and skills related to continuous engineering practices varies among these specialties: while software developers use tooling to achieve a high degree of automation in their daily work, data scientist and, to a lesser extend, data engineers have a less structured way of working. Many data scientists are not aware of tools like version control systems that can track efficiently changes in the training code and the data used to perform the experiments, or ticketing systems that enables them to track progress from the feature implementation to requirements. Instead, they rely on bespoke solutions that might not be appropriate when practiced in the development process of safety critical systems.

To incorporate ML features in software development, several techniques have been proposed.
In this work, we build on Continuous Delivery for ML \cite{cd4ml} and ML Model Cards \cite{mitchell2019model}. These techniques are briefly introduced in the following. 

\subsubsection{Continuous delivery for ML}

\begin{figure}[htb]
    \centering
    \includegraphics[width=\textwidth]{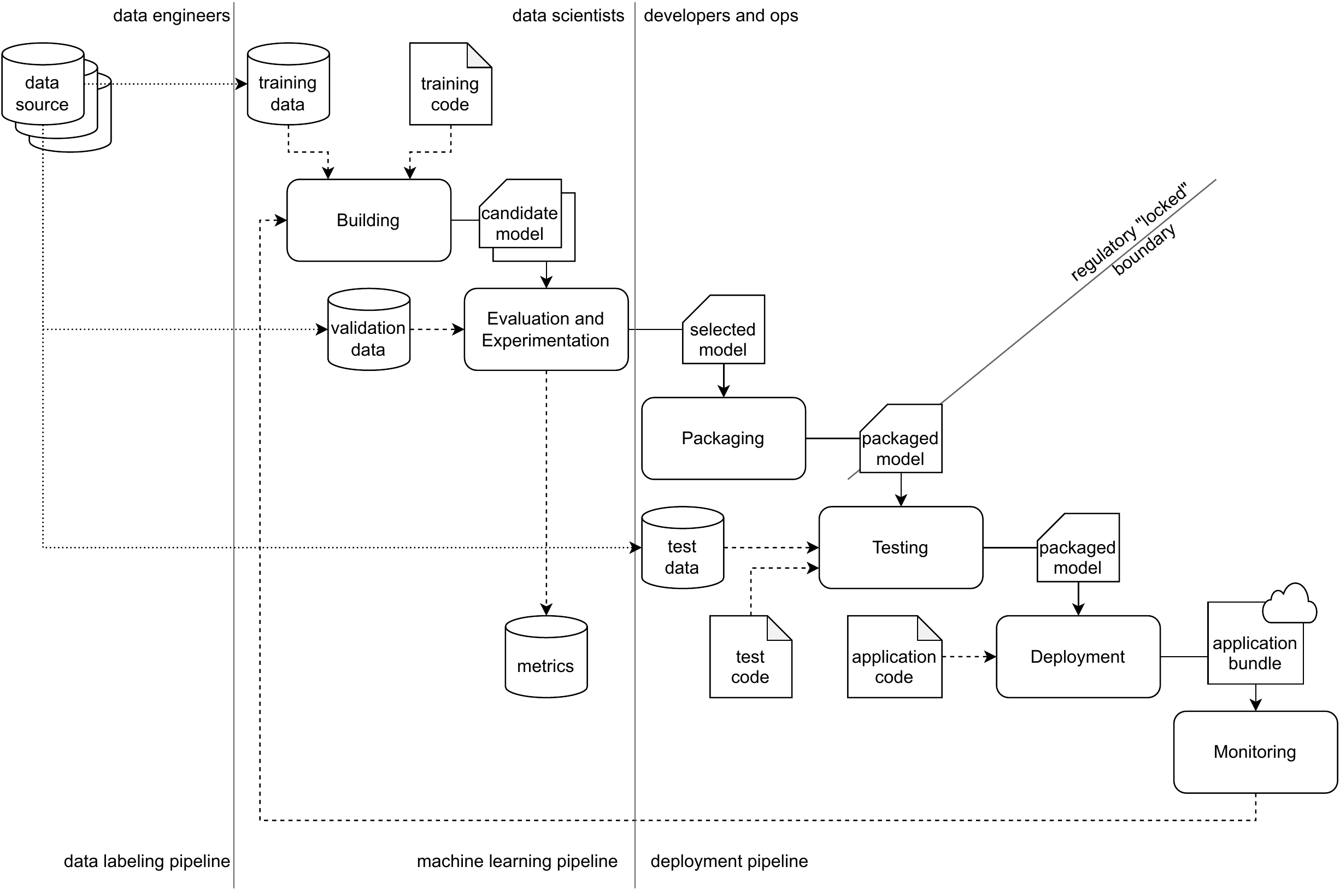}
    \caption{CD4ML pipelines and artifacts. Figure adapted from \cite{oravizio}.}
    \label{fig:cd4ml}
\end{figure}

Probably the best-known MLOps implementation, Continuous Delivery for Machine Learning (CD4ML) \cite{cd4ml} by ThoughtWorks, aims at automating  the ML application life-cycle in an end-to-end fashion (Fig. \ref{fig:cd4ml}). In CD4ML, a cross-functional team produces applications based on code, data, and models in small and safe increments that can be reproduced and reliably released at any time, in short adaptation cycles. Three distinct steps are included: (i) identify suitable data sources and prepare the data for training, (ii) experiment with different models to find the best performing candidate, and (iii) deploy and use the selected model in production as a part of a bigger software system.  

CD4ML is not the only solution that aims to prevent the accumulation of hidden technical dept in machine learning applications \cite{ml-tech-dept}. While other solutions are heavily optimised for a particular cloud infrastructure \cite{aws-mlops,google-mlops}, an ML software stack implementation \cite{tensorflow-extended}, or conceptual characteristics of the approach like MLOps \cite{john2021towards}, the CD4ML model has the advantage that it can be used as a reference model in any application domain. Furthermore, it has the necessary phases documented at the appropriate level, and the implementation is based on open source components. Hence, it serves as a solid foundation for experimentation, and to translate the results to other other MLOps pipelines. 

\subsubsection{ML model cards}

ML model cards is a framework for transparent communication of information that facilitates the correct utilisation of machine learning models \cite{modelcards}. Intended as documentation that accompanies a trained machine learning model, the model card contains information about intended use case, benchmark evaluation in a variety of relevant conditions, such as demographics or geographic location, or any other information that the creators considers suitable for the proper use of the trained model. Although the original proposal presented the model card as a visual representation, recent developments\footnote{https://github.com/tensorflow/model-card-toolkit} propose a programmatic mechanism to generate the model cards, and a machine readable serialization\footnote{\href{https://github.com/tensorflow/model-card-toolkit/blob/master/model_card_toolkit/schema/v0.0.2/model_card.schema.json}{TensorFlow model card schema}} that facilitates the consumption of model cards by scripts in ML pipelines.

While the machine readable serialization of the model card is a step in the good direction, the information included in the card is limited to specific audiences, like data scientists or machine learning engineers. To fulfil the model cards potential, the information should be extended to include the needs of other stakeholders. The model card metadata becomes a model from which various views targeted at different stakeholders can be derived, in a similar fashion as various software architecture views can be derived from a single software model. The approach will turn the model card into a must have artifact that conveys not only the information about the model's intended use or performance, but also what other activities have been conducted in relation with the model (e.g. regulatory risk management), or metadata about data sets that facilitate downstream testing \cite{oravizio}.

\subsection{Medical regulatory landscape}
%Landscape
The manufacturing of medical devices is strictly controlled by authorities, and manufacturers must conform to the region's regulatory requirements in which a medical device is being marketed for use. For this reason, medical software systems must also be developed according to the requirements of the target area. For example, the development is regulated by Medical Device Regulation (MDR) \cite{mdr} and In Vitro Diagnostics Regulation (IVDR) \cite{ivdr} in the EU region and by Federal Food, Drug, and Cosmetic Act (FD\&C Act) \cite{fdc_act} in the USA. 

\subsubsection{Design control}

The regulatory framework aims to ensure that a medical device is safe to use and clinically effective for its intended medical purpose. In practice, there are certain mandatory processes that include control mechanisms for the whole software lifecycle, including design, development, and manufacturing of the product. The regulatory requirements related to these process phases are generally referred to as the Design Controls \cite{fda-design-control}. 

\begin{figure}[htb]
    \centering
    \includegraphics[width=0.7\textwidth]{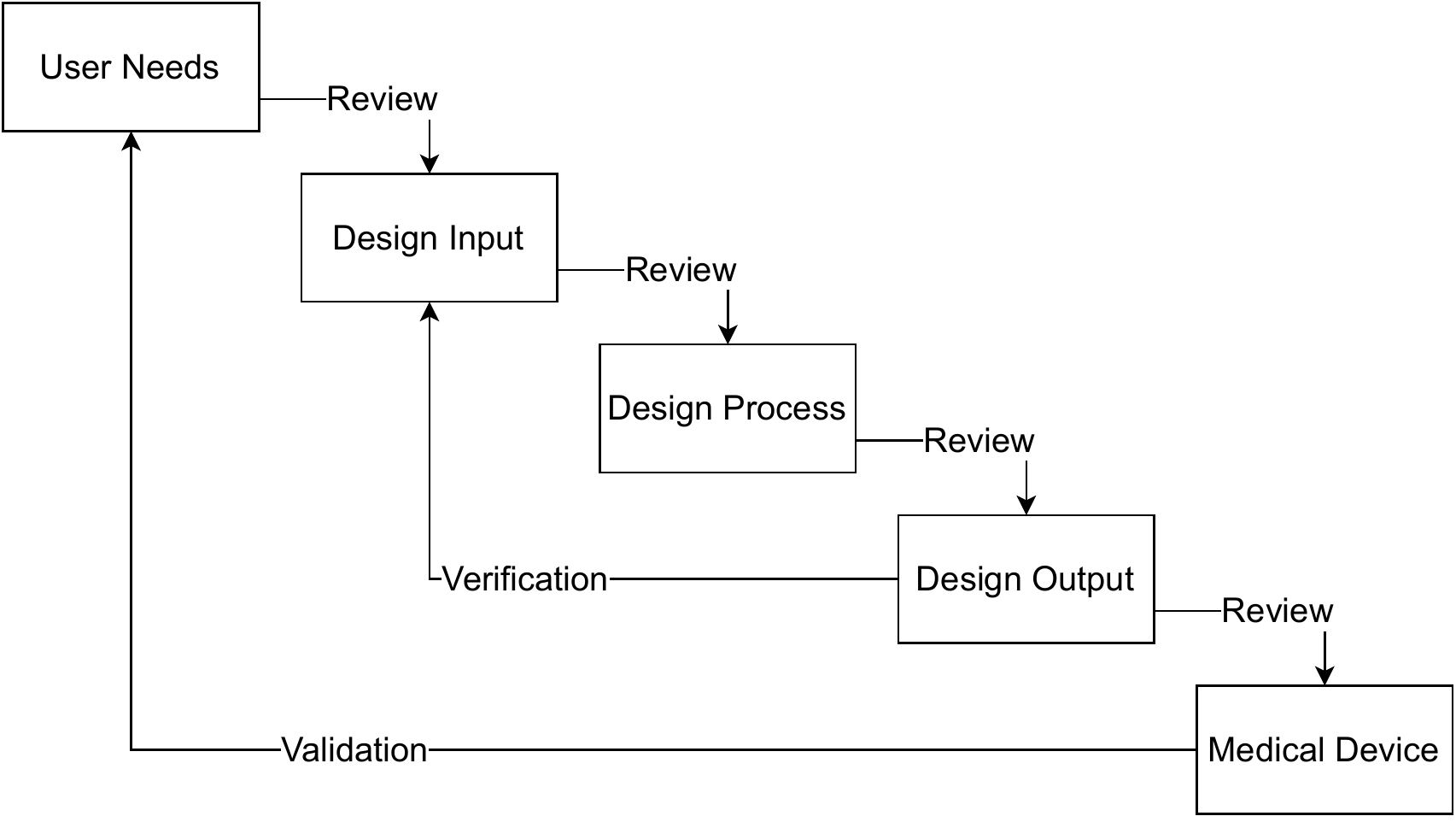}
    \caption{Design control process for medical devices. Figure adapted from \cite{fda-design-control}.}
    \label{fig:design-control}
\end{figure}

The purpose of the Design Control process (depicted in Fig. \ref{fig:design-control}), is to promote a well-designed development process that includes traceability between process inputs and outputs at different stages of the process. Starting from \textit{user needs} converted into \textit{design inputs}, continuing with the \textit{design process} that transforms the inputs into \textit{design outputs}, and finally forming the resulting \textit{medical device}. In addition, the reviews during each step of the process verifies and validates that the requirements are met by the implementation, reducing the possibility of design and implementation defects. In general, these regulatory boundaries are addressed in software development with dedicated medical software lifecycle management tools, such as Polarion\footnote{https://polarion.plm.automation.siemens.com/products/polarion-alm}.

\begin{figure}
\centering
\includegraphics[width=0.95\textwidth]{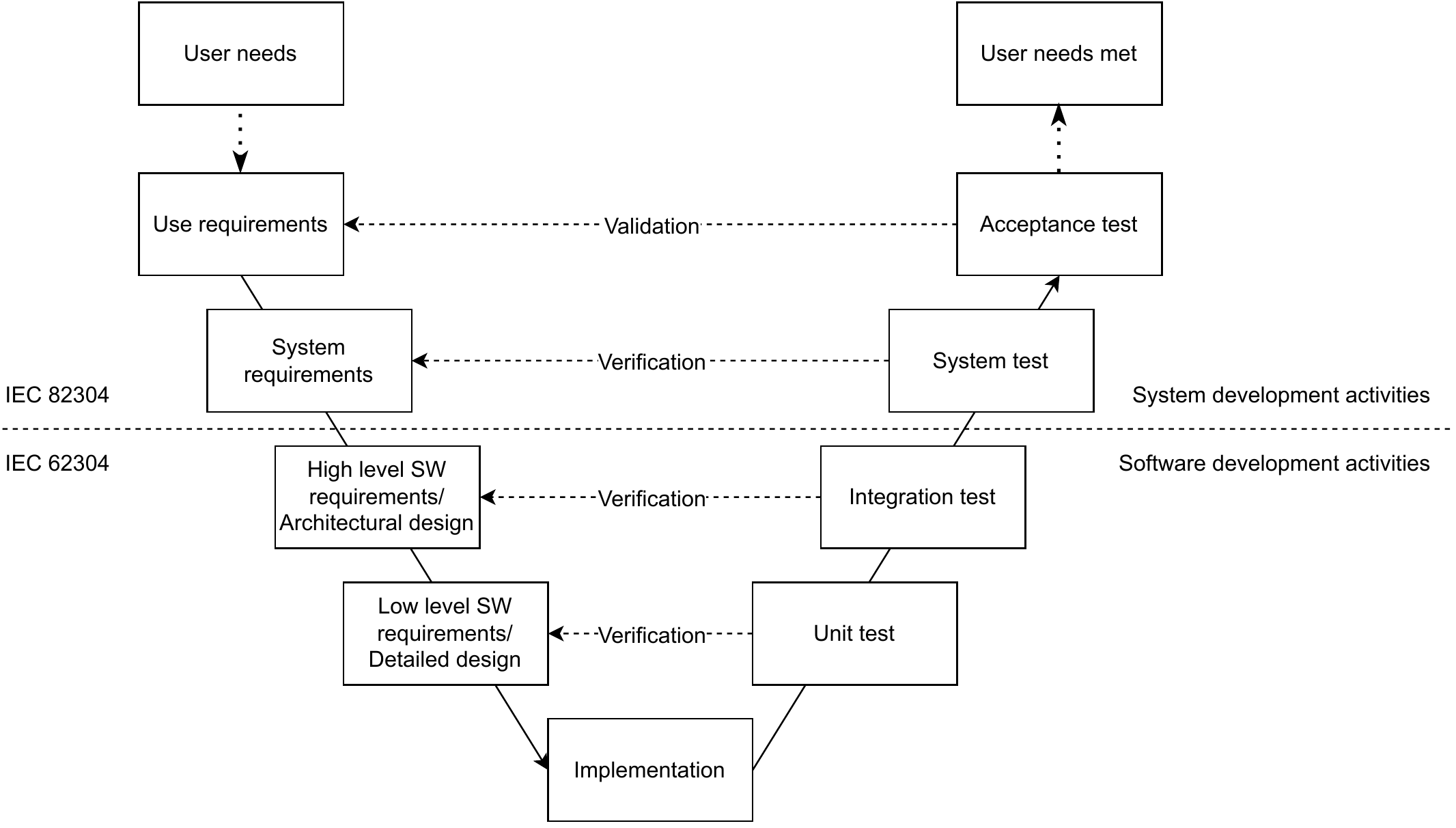}
\caption{System and software development design control activities.} 
\label{fig:sdlc-activities}
\end{figure}

For software medical devices the design control is implemented in two layers, depicted in Fig. \ref{fig:sdlc-activities}: the product and system development activities (IEC 82304 \cite{iec82304}), and the software development activities (IEC 62304 \cite{iec62304}). At the product level, the identified user needs are converted to system requirements that serve as design inputs for the the software development process. During software development, the system requirements are transformed into high level software requirements that cover the software system and architectural concerns. Later on, the high level software requirements are further distilled into low level software requirements that serve as design input for implementation.

The architectural design activity defines the major structural components of the software, known as \textit{software items}. It identifies their key responsibilities, their externally visible properties, and the relationship among them. The resulting software architecture artifact ensures the correct implementation of the software requirements, and is complete when all software requirements can be implemented by the identified software items. The architectural decisions are extremely important for implementing effective risk control measures. The proper understanding and accurate documentation of software items behaviour is essential for ensuring that the software system is safe. Detailed design activities refine the identified software items during architecture design into smaller software items. When a software item is not decomposed further it is called \textit{software unit}. In the end, the manufacturer is responsible for the granularity of the software decomposition, and should ensure that the activity performed to the appropriate detail to allow a safe and effective implementation.

The resulting code, test cases and various other artifacts, such as architecture and detailed module design documentation, created during the software development activities, serve as the design outputs. The review of the artifacts and the test result provide an effective verification procedure at unit, integration and system level. The acceptance tests together with the result reports of clinical trials serve as the validation procedure. All these procedures ensure that the proper design controls have been applied during development, resulting in a medical product that meets the user needs

\subsubsection{Design control for ML: the missing parts}
\label{sec:objectives}
Although the design control process for medical software is otherwise well-defined, the regulatory frameworks of the major market areas, such as the EU and USA, do not currently explicitly address the requirements related to AI/ML technologies. While AI/ML-based medical devices have great potential to improve care for individual patients, they carry some unique risks. For example, certain ML systems may be designed to learn and optimize their functionality in real-time. As a result, the user-facing risk profile of these systems may change over time which is an incompatible idea compared to general design control practices. Evidently, the current lack of precise requirements for AI/ML is a severe shortcoming in the legislation as it creates particular challenges and uncertainties for the manufacturers on how to prove AI/ML device safety and efficiency when seeking device approvals. 

In general, the AI/ML devices must comply with requirements applicable to all medical device software. Therefore, manufacturers are required to manage the risks related to their products throughout the devices' whole life cycle \cite{mdr, granlund_cybersec}. In practice, medical software risk management activities are implemented according to the requirements of ISO 14971 \cite{iso14971} and IEC 62304 \cite{iec62304}. However, the risk management activities described in detail in the standards mentioned assume that the device's functionality remains the same as during the product development phase, even after deployment, and does not change over time. As a result, it may be difficult for the AI/ML manufacturer to prove the effectiveness of the implemented risk management activities against the current requirements.  

Furthermore, specific AI/ML models can be complex and data-intensive, so they may be complicated to understand fully. As a result, it may not be easy to assess how the model has reached a decision. In addition, the quality of data plays a significant role and may contain biases not visible for a human auditor. Also, specific challenges related to model training, such as overfitting and underfitting, must be considered when validating the system's clinical efficiency and safety. 

There are currently certain ongoing efforts to address the problem of missing AI/ML regulatory guidance. For example, in the EU, the theme "Artificial Intelligence under MDR/IVDR framework" will be addressed within the forthcoming guidance document by the Medical Device Coordination Group (MDCG) \cite{mdcg-ongoing}. Furthermore, in the USA, the US Food and Drug Administration (FDA) has released an action plan document "Artificial Intelligence/Machine Learning (AI/ML)-Based Software as a Medical Device (SaMD) Action Plan" \cite{fda-action-plan}. While specific and binding regulatory requirements are still under development, the Interest Group of the Notified Bodies for Medical Devices in Germany (IG-NB) has created perhaps the most detailed and concrete guideline available currently, "Guideline for Artificial Intelligence in Medical Devices" \cite{ig-nb}. The IG-NB guideline can be used as a basis to gain an understanding of the expectation level of the notified bodies related to AI/ML products. 

\section{Proposed Solution}
\label{sec:proposed_solution}

The proposed design control process for CD4ML
pipelines aims at formalizing software development so that it be used for QMS purposes. This is achieved by using pull requests as basis for reviews that forms the design control, and using model cards metadata as the design output artifact that serves also as an audit trail for regulatory activities such as clinical validation and risk management. 

\subsection{Applicability considerations and background}

The proposed solution is used in an environment where modern software engineering methodologies (e.g. continuous software engineering), tools (e.g. Git) and practices (e.g. change management with Git pull requests) are used. These facilites are used as enablers to implement traceability and design control activities for medical certified systems are implemented.

\subsubsection{Traceability}

\begin{figure}
    \centering
    \includegraphics[width=0.7\textwidth]{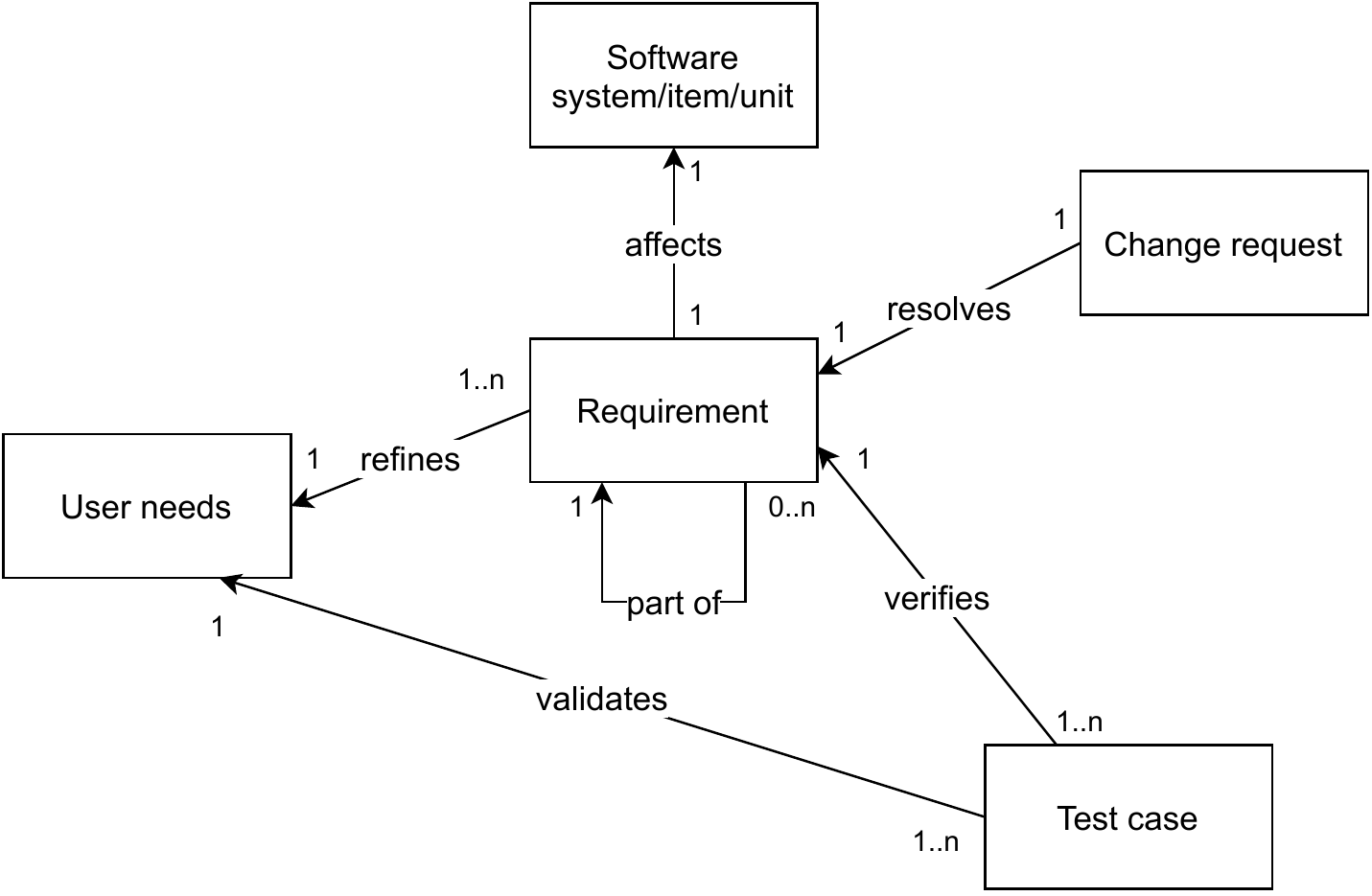}
    \caption{Traceability information model.}
    \label{fig:traceability-model}
\end{figure}

Vendors like GitHub and GitLab offer a wide range of capabilities, in addition to the original version control system enabled by Git\footnote{https://git-scm.com}. For example, \textit{issue} are used during implementation to track tasks or bugs, but also to plan future work by collecting ideas and feedback. Linking issues allows building a graph that enables the users of these systems to navigate the links to understand how the individual items of work relate to each other. These capabilities can be used to build an effective traceability system that is familiar to developers working in agile projects \cite{stirbu2021traceability}.

A traceability information model, described in Fig \ref{fig:traceability-model}, starts from \textit{user needs} that are decomposed recursively into system and software \textit{requirements}. Requirements are resolved by the \textit{change requests} that are verified by \textit{test cases}. Each requirement is mapped to the corresponding \textit{software system, item or unit} resulted from the software decomposition activity.

\subsubsection{Design control with pull requests}

\begin{figure}
    \centering
    \includegraphics[width=\textwidth]{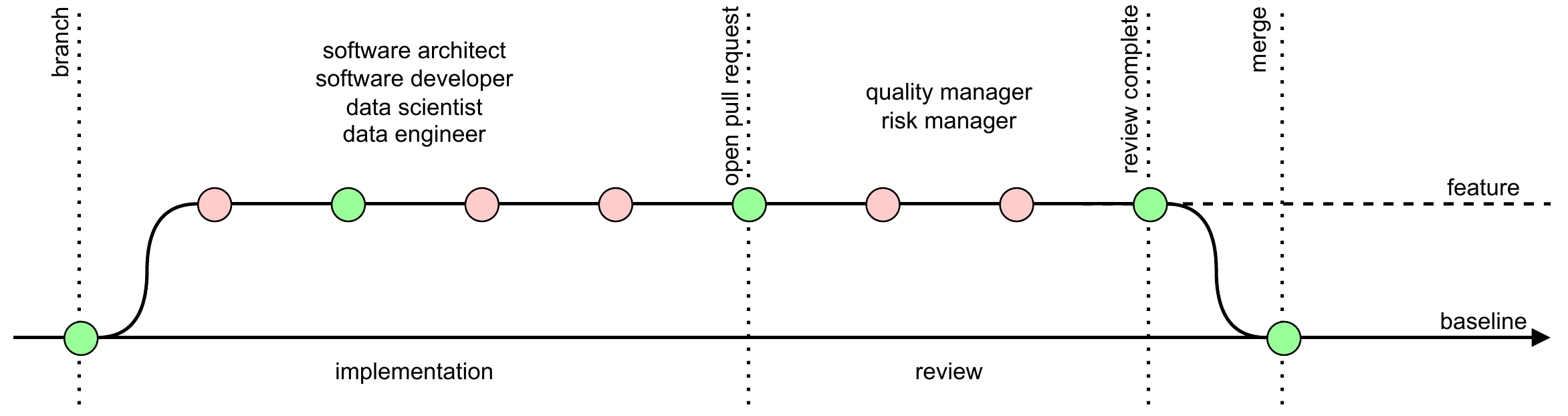}
    \caption{Design control with pull request.}
    \label{fig:pull-request}
\end{figure}

Using pull requests has become the main practice used by development teams to introduce code changes into the code baseline. As depicted in Fig \ref{fig:pull-request}, the process starts with creating a new branch that will contain the changes needed to resolve a requirement. During the development phase, software architects, software developers, data engineers and  data scientists create new artifacts that are added to the branch as commits. When they consider that the implementation is ready they open a pull request to signal the rest of the team that the review can start. During the review, the team members that conduct code review activities are joined by other specialists that perform regulatory activities such as risk management. The new functionality can be merged into the baseline when the review is complete and all required activities have been completed successfully. For each commit the Git vendor system performs a set of automated checks that enables team members to focus on important activities that require human supervision, leaving many compliance repetitive tasks to tooling. 

The pull request practice provides a team wide venue for conducting sophisticated reviews that are not  limited to code only \cite{stirbu2018towards}. Team specialists that are not typically involved in software development are able to perform their activities during the pull request review, at the granularity of each introduced change. As all team members work towards a common goal during each iteration, the baseline is always up to date from regulatory perspective \cite{calm-compliance}.

From a design control perspective, the requirement constitutes the \textit{design input}, the set of artifacts committed to the branch during the iteration, seen as \textit{design development}, represent the \textit{design output}, and the pull request review correspond to \textit{review}, and \textit{verification or validation}. Overall the pull request is an effective design control mechanism for feature-branching development model. Besides the review, the pull request serves also as a traceability audit trail artifact.

%\vlad{clarify what change request is in practice for feature-branch or trunk-based development}
Although we have described the design control implementation using pull request, a similar result can be obtained for teams using the trunk-based development model, in which all changes are committed into the baseline. The team practices can include additional conventions, such as adding the requirement identifier to the commit message, that serve as an effective traceability mechanism for relating commits and corresponding merge reviews to a single requirement.

% \todo{Describe the pull request as a org wide venue for having reviews, not limited to code.}

\subsection{Integrating ML models into certified medical systems}
\label{sec:integration}

Machine learning models must be integrated into the larger software system before they can be used. In general, the selected model is integrated using one of the following integration patterns: include the model in the application code as a \textit{library}, run the model as a \textit{service}, or dynamically load the \textit{model as data} at runtime from a remote repository.

The use of machine learning technologies in medical systems is governed by domain specific restrictions, which are typically introduced in the US and EU implementation guidelines. To properly handle the risk that is inherent with the non-deterministic behaviour of machine learning, the models must be in a \textit{locked} state, which means that it is not allowed to change the models once they are deployed in production. Considering this restriction, we can assume that the practical ways of integrating machine learning models into medical systems is either as a library in the software item in which the model will be used, or as a machine learning service that exposes its functionality to other software components over the network. In this work, we assume that this library or service must be associated with the corresponding chain of provenance. 

\begin{figure}
    \centering
    \includegraphics[width=0.5\textwidth]{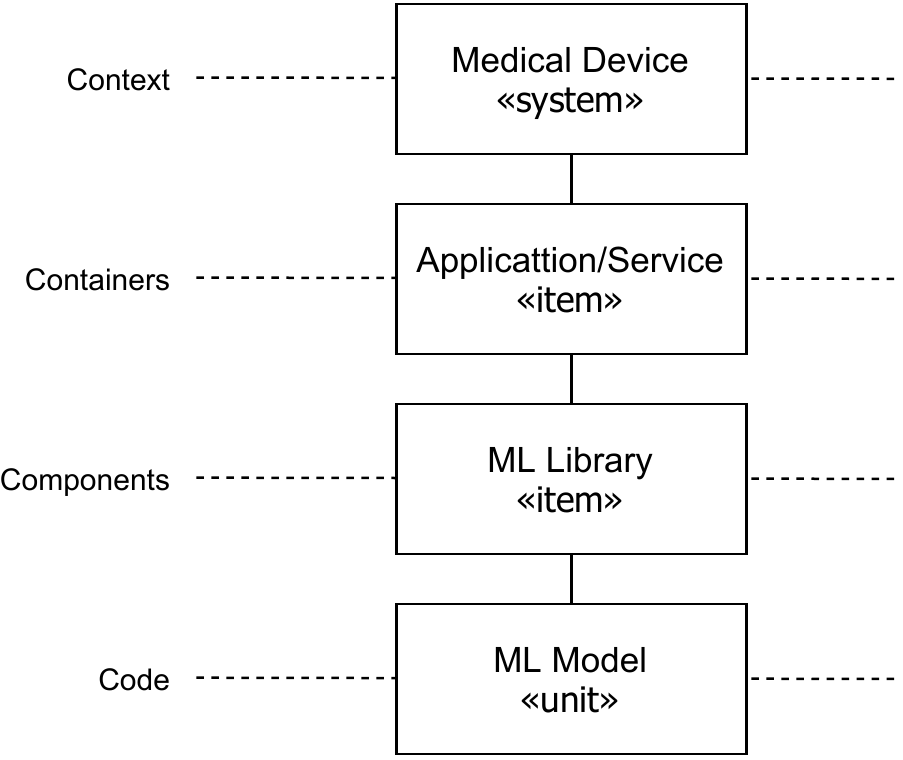}
    \caption{Example software decomposition of a medical software device, mapped to the four abstraction layers introduced by the C4 model.}
    \label{fig:decomposition}
\end{figure}

Looking from the perspective of the architectural and detail design regulatory activities, in which the medical system is decomposed into items and units, we can observe that the ML model developed by the data scientists is integrated into the medical system as a software unit. The selected model is first packaged as a library, and then is integrated into an application or service, to reflect the integration options listed above. The software decomposition and the model integration, mapped on the four abstraction layers defined in C4\footnote{https://c4model.com} model for visualising software architecture, is depicted into Fig \ref{fig:decomposition}.

\subsection{Continuous ML design control}

Pull requests have been proposed to act as a design control for regulated software development \cite{stirbu2018towards} in the spirit of DevOps. Since the whole CD4ML pipeline relies on software, we expand the same concept to data engineers and data scientists tasks, related to the pipeline. This effectively means expanding the types of artifacts that are stored and versioned to data. In other words, pull requests become the design control for all changes in the development.

\subsubsection{Activities performed during implementation}

The model implementation starts with data engineers loading and labeling data from different sources and transforming the input into two datasets that are used for training and testing. The training dataset is used to develop a model and to perform a series of experiments that determines which model candidate perform best. A model card metadata document is created once the best performing model candidate is selected. The medatada document, which contains the model's documentation and the associated artifacts that enable quality control when the model, is integrated in the medical system.

\begin{figure}
    \centering
    \includegraphics[width=\textwidth]{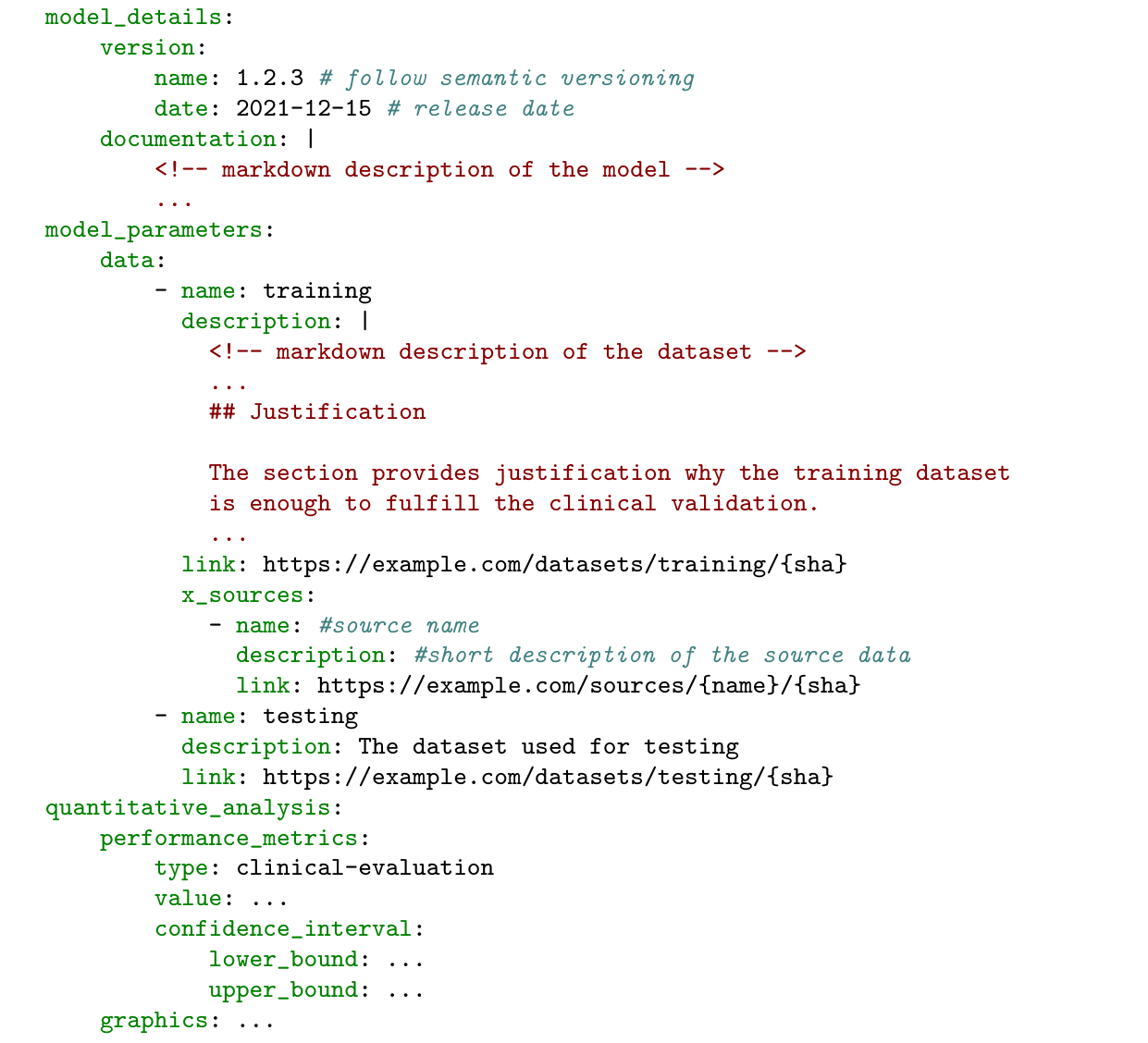}
    \caption{Expanded model card metadata.}
    \label{listing:documentation}
\end{figure}

The model card document, presented in Fig. \ref{listing:documentation}, contains a thorough description of the model and its usage (preferably in Markdown format), versioning information, and information about the datasets used for training and testing, together with the information about the data sources (e.g. \texttt{x\_sources}) used to create the datasets. The information about the dataset is used in the pipelines during the packaging and integration steps. Additionally, metadata includes the information collected during the clinical evaluation of the selected model. The content of the metadata document is created or updated by the data scientists and data engineers that contributed to the development of the corresponding model iteration.

The modelcard metadata document together with the model code and the test dataset constitutes the design output. To perform an effective continuous design control process, the artifacts must be reviewed and ensure that they fulfil the requirements using the team established change request procedures that leverage the pull request. From the CD4ML perspective, these activities corresponds to the \textit{building} and \textit{evaluation and experimentation} phases. Additionally, when the selected model is merged into the mainline, it is packaged and published into a model repository from where it can be used downstream.

\subsubsection{Activities performed during integration}

The ML model is integrated into the medical system either directly into an application that uses it locally, or as a service that is invoked over the network by other components in the system. Depending on the integration option, the manufacturer must ensure with appropriate design control procedures that the ML model is used as intended by its creators. The integration testing must include a test suite that ensures that no deviations are introduced in the use of the ML model. The test suite makes use of the test dataset provided in the model card metadata document. The result of the integration test represent the clinical performance evaluation and should be included in the \texttt{quantitative\_analysis} section of the model card metadata document. These activities are part of the regulatory software development lifecycle \textit{integration}'s activities and the CD4ML's \textit{testing} and \textit{deployment} steps.

\subsubsection{Activities performed during release}

When releasing the medical system the model card metadata should be used to automatically generate the the clinical validation report in a format appropriate for regulatory purpose (e.g. auditors, certification). At this point, the resulting documents can be published to an external document management system.

\subsection{Risk management with model cards}

When developing the machine learning model, the team needs to perform risk analysis throughout development lifecycle. From a regulatory perspective, a medical system lifecycle consist of two major parts: \textit{pre-market} defining the period during which the product is initially developed, and \textit{post-market} defining the period in which the products is used by intended users, during which the product is maintained. From a continuous engineering perspective, where the product is continuously improved the pre/post market separation is not that strict as development and maintenance activities are routinely performed during a single increment. As machine learning models in medical systems are in a regulatory \textit{locked} state, all changes and risk management activities are allowed only in pre-market phase, the post market being reserved for collecting user's feedback or for handling the faults and anomalies detected by the monitoring systems.

\subsubsection{Risk management activities performed pre-market}

Algorithms are an essential ingredient of machine learning. The risks inherent to algorithm design propagate to medical machine learning applications due to their increased complexity, lack of transparency, inappropriate use, or weak governance. Algorithmic risk can be split into three categories \cite{algorithmic-risk}: input data, algorithm design and output decisions. Flaws in input data such as biases in the data used for training, the quality of the data can lead to mismatches between the data used for training and the data used during normal use. Output decisions flaws relate to incorrect interpretation or use of the output. Algorithm design flaws can be expanded in human biases – cognitive biases of model developers and users can lead to flawed output, technical flaws – lack of technical rigour or conceptual soundness during development, training, testing and validation, usage flaws – incorrect implementation or integration in operations can led to inappropriate decision- making, or security flaws – threat actors can manipulate the inputs to produce deliberate flawed outputs.

The machine learning related risk analysis activities should document their findings. The data labeling process needs to be accompanied by the justification on why the data sources used for building the training data is enough to fulfill the clinical validation. The documentation needs to be added to the training dataset section in the model card metadata document. The identified risks and the possible mitigation strategies needs to be documented as requirements and serve as input documents for development. Similarly, limitations, trade-offs and ethical considerations need to be documented in the appropriate section of the metadata document (e.g. \texttt{considerations}).

\subsubsection{Risk management activities performed post-market}

As the model in a deployed medical system is in locked state, any corrective actions for mitigating the anomalies and deviations detected by the users or by monitoring systems serve as input for a future development iteration. \new{For example, the monitoring infrastructure that has the ability to detect deviations in the average accuracy and confidence of a deployed model can lead to the discovery of new input data, that may relate to model drift, or changes in the underlying relationships between input and output data, that may reveal the possibility for concept drift.} These events have to be documented using the \new{regulatory required} user feedback \new{procedures \cite{iso13485},} or as software bugs and converted into new requirements, \new{following a specific root cause analysis investigation activity}. \new{From the CD4ML perspective, the ML model corrective activities are part of the feedback loop that connects the monitoring stage to the building stage.}

\subsection{Revised design control process}

Our approach for implementing continuous design control aligns the activities conducted while developing machine learning applications, exemplified with the CD4ML pipeline, with the rigour expected when developing certified software medical systems. The alignment of these activities is two fold: to harmonise the terminology and to identify the artifacts that serve as audit trails that the machine learning development has been implemented in line with the requirements of IEC 62304, which governs the development of software used in medical products.

\begin{figure}
    \centering
    \includegraphics[width=\textwidth]{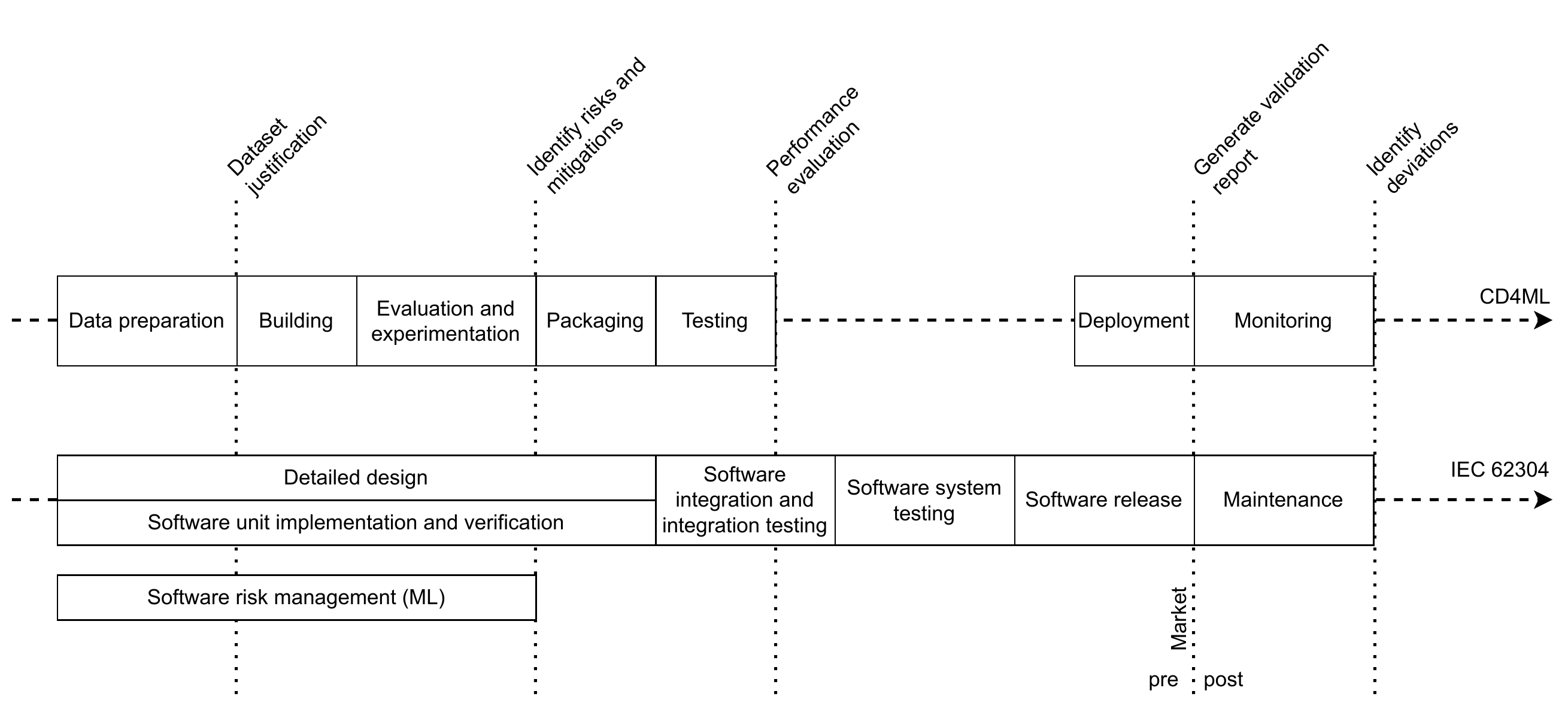}
    \caption{Design control process.}
    \label{fig:revisited}
\end{figure}

In practical terms, the pull request constitutes the design control mechanism that ensure that the evolution of the system is systematically reviewed and that the code baseline is always up to date from a regulatory perspective. Besides aligning the development and the regulatory activities, we identified the model card metadata as an ideal candidate for documenting not only the model but also the regulatory specific activities performed during model's development, such as dataset justification or clinical performance evaluation, among other pre-market risk management activities. The model card metadata document, together with the model code, serves effectively as the design output artifact. Being machine readable, the model card metadata can be used in pipelines to generate automatically additional documents intended for end users (e.g. the model card), and regulatory authorities (e.g. clinical validation report). Post-market monitoring and maintenance activities that identify deviations from the expected model behaviour are identified and captured as bug reports or feedback and fed into the team backlog as requirements. The terminology and development phase harmonisation together with the design output artifacts are described in Fig. \ref{fig:revisited}.

\section{Case Study: Oravizio Process Revised}

In our previous work, we have introduced Oravizio\footnote{https://oraviz.io/}, CE certified medical software for assessing the risks of joint replacement surgeries \cite{oravizio}. We use this system to demonstrate how to apply continuous design control for ML in certified medical systems. The work is a concept prototype in its nature; it builds on experiences from an industry system, but the proposed implementation has not been deployed to industrial use.

Our previous work with Oravizio has introduced a continuous training pipeline. The pipeline allows to overcome regulatory constraints associated with Oravizio ML model training and to simultaneously achieve automation goals associated with MLOps \cite{oravizio}. In addition, the pipeline addresses the medical device software design control requirements by design. 

\begin{figure}[htb]
    \centering
    \includegraphics[width=\textwidth]{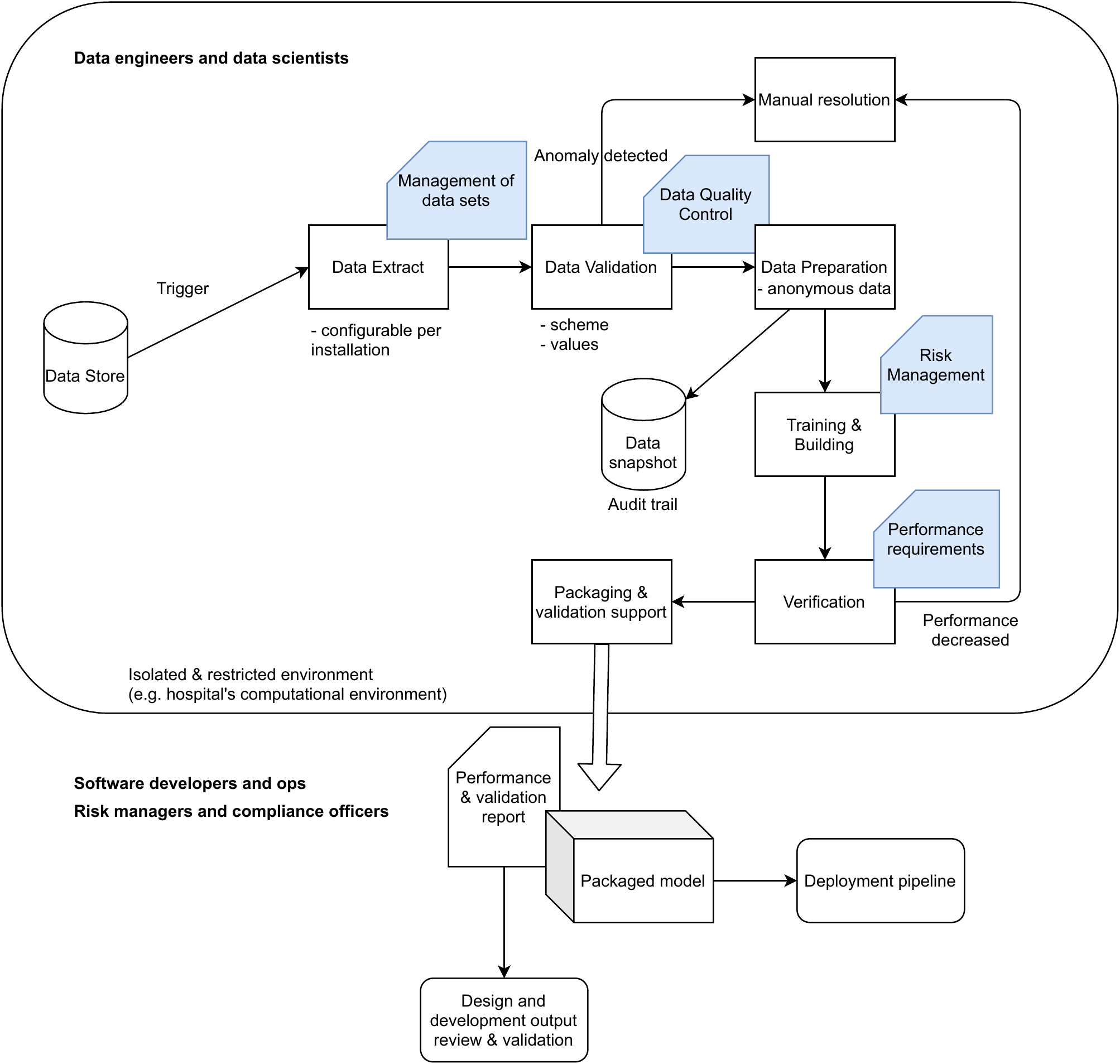}
    \caption{Continuous training pipeline with model cards. Arrows indicate data flows.}
    \label{fig:continuous-pipeline}
\end{figure}

For this paper, we have revised the pipeline by extending it with a carefully selected set of model card documents, to demonstrate the proposed solution. %presented in Section \ref{sec:proposed_solution}. 
Fig \ref{fig:continuous-pipeline} presents the pipeline of \cite{oravizio}, with the proposed model card extensions marked with a darker color. 

%Subsection "what is today"
\subsection{Continuous training with MLOps pipeline}
The data used to train and re-train the Oravizio ML models is generated within the clinical processes of a collaborating partner hospital, and, by design, Oravizio does not generate corresponding data in production use. Because of the sensitive nature of data, access to the hospital's computational environment is strictly restricted. As a result, the continuous training pipeline was designed to operate inside the clinical partner's controlled environment and fetch new data from the data store based on pre-defined triggers. Furthermore, as the models are deterministic by nature, their \new{technical} performance can be validated \new{according to the principles of clinical evaluation} with test data in \new{a restricted} environment without the need to do the \new{intended medical use specific} validation in the final production environment. Finally, Oravizio was designed to be deployed in a production environment with its ML model in a "locked" state for regulatory reasons. In practice, Oravizio's models are trained during the development phase, and their ability to improve the outcome on the fly is disabled in production use. Despite this limitation, the pipeline enables the laborious task of re-training to be automated. 

In addition, the fact that the development team does not need access to the restricted environment beyond the pipeline's installation and maintenance is a valuable design feature. Furthermore, the pipeline can automatically generate the required documentation needed to assess the model performance in the clinical performance evaluation. All created artifacts are delivered to the development team from the isolated environment. 

%subsection "how this affects"
\subsection{Design control documentation with model cards}

Even if the original version of the pipeline in \cite{oravizio} contains the required design control documentation, which can be generated automatically, the documentation format has not been based on any generally known standard. The reason is that at the time of implementing the pipeline, no such format was available \cite{mitchell2019model}. In addition, the selected document templates have been similar to more traditional quality management system types of records targeted to regulatory stakeholders and without the technical ability to be serialized. To address these challenges and further support continuous design control for ML, we expanded the pipeline with the model cards tailored to address the design control documentation requirements. 

\subsubsection{Data set management}
To ensure the required performance of the models, the selected data set must be representative of the target population. In Oravizio's case, data sources could differ per installation, and, as a result, the procedure for data extraction must be configurable. In addition, there are many related documentation requirements, which can be documented with a model card. Firstly, the allowed data sources must be listed with the specific requirements and descriptions for a data source. Secondly, the data inclusion and exclusion criteria must be defined with the procedure for invalid data management. Thirdly, data protection policy needs to be defined with clear instructions on how to ensure data protection at later stages of data processing. Finally, potential biases in the data must be reflected and the selections made justified accordingly. 

\subsubsection{Data quality control}
As part of data quality control for a system using supervised learning, such as Oravizio, the labeling and label verification procedures are essential. Moreover, when utilizing automation in re-learning, the correctness of labeling needs to be constantly monitored. It is possible to use different pre-defined schemas and boundaries to validate the data quality, both in terms of format and content. As model cards are already in a machine-readable format, they can be used simultaneously as a documentary and a validating artifact. If the rate of error in the data validation rises above the acceptable limit, the re-learning cannot continue automatically, and the anomaly must be resolved manually. 

\subsubsection{Risk management}
In conjunction with clinical evaluation, effective risk management provides a practical tool for the manufacturer to prove the safety and clinical efficiency of the device. As the ML models are a central part of Oravizio, the models have a considerable impact on the device's risk profile. As a result, the potential risks related to the models need to be carefully addressed. 

When utilizing model cards, the identified model-related hazards and corresponding risk mitigations are documented on the model card document. It is worth noticing that certain risks, both in terms of patient safety and data security, are unique to AI/ML systems. These risks include, for instance, model drift, the drift in data distribution, and risks related to continuous learning systems. In addition, also risks related to adversarial attacks need to be considered.  

\subsubsection{Performance requirements}
According to the regulatory requirements, manufacturers of medical devices must document the intended purpose of their device, including specification of indications, contraindications, patient target groups, and operating parameters and limitations \cite{mdr}. In addition, performance characteristics, accuracy, and the limits of accuracy, precision, and analytical performance must be addressed if applicable. 

Many of the above requirements apply directly to Oravizio's ML models. Moreover, they can be conveniently documented within a model card. 

\section{Discussion}

In this section, we provide an extended discussion regarding the proposed approach, highlighting how it brings benefits to the involved stakeholders. In addition, we discuss limitations of this work.

\subsection{Aligning ML development and regulatory practices}

The use of ML technologies in certified medical devices is an emerging trend in a notoriously conservative industry. Consequently, the regulatory practice is not as established as the practices in the development of traditional software medical devices. 

Our proposal brings together the ML application lifecycle, demonstrated using CD4ML, and the medical device software lifecycle process standard IEC 62304. The result lowers the cognitive barriers between the machine learning model developers, such as data engineers and data scientists, and regulatory practitioners, allowing them to work together to effectively develop medical devices that include machine learning technologies.

\subsubsection{Avoiding common ML system design problems}

\begin{table}[p]
\begin{center}
\begin{minipage}{\textwidth}
\caption{Mitigate common ML system design problems \cite{ml-tech-dept} with continuous design control}\label{table:mitigation}%
\begin{tabular*}{\textwidth}{p{28mm} p{40mm} p{40mm}}
\toprule
ML system design problems & Mitigation strategy & Model card document role\\
\toprule
Configuration: \textit{\mbox{Do we know} all configuration options and their effects?} & Configuration management is a prime concern in the SDLC of medical systems \cite{iec62304}. The existing practices established by the medical device manufacturer for \textit{configuration} and \textit{software risk management} can be expanded to cover the configurations of the ML model integrated into the medical system. & Contains the ML model parameters and their valid ranges for the intended use. \\
\midrule
Data collection \& Feature extraction: \textit{Do we know if the input data and features developed are enough for the intended use?} & The review includes an analysis of the selected data sources and the methods used for feature extraction. & The model card is extended to capture the source for each data set. Together with the justification section included description, it provides the evidence of needed to fulfill the clinical evaluation.  \\
\midrule
%Feature \mbox{extraction} & & \\
%\midrule
Data verification & The relevant data set is used to verify the integration for each change request. & The verification data set is included in the model card, and should be used by the relevant MLOps integration stages. \\
\midrule
Resource management: \textit{Do we know that the needed hardware and software resources are available to ensure the model performs correctly?} & The quality management system, implemented by the manufacturer to fulfill regulatory requirements \cite{iso13485}, ensures that the needed resources for the proper functioning of the medical product are allocated. & The model card contains the information about the special resources that are needed to run the model. The information should be used downstream when planning the resource allocation.  \\
\midrule
Serving infrastructure: \textit{Do we know that the model is integrated and works properly?} & ML models are packaged and integrated into medical systems as \textit{libraries} (see Section \ref{sec:integration}). As such, their functionality is exposed via an API and changes to the ML model are contained. & - \\
\midrule
Monitoring: \textit{Can we detect deviations while the model is used with production data?} & Medical device manufacturers must establish post-market surveillance program \cite{iso13485, iec62304}. The program should cover the monitoring the ML components in use. & The quantitative data included in the model card document serves as input for monitoring components that detect deviations. \\
\botrule
\end{tabular*}
% \footnotetext{Source: This is an example of table footnote. This is an example of table footnote.}
% \footnotetext[1]{Example for a first table footnote. This is an example of table footnote.}
% \footnotetext[2]{Example for a second table footnote. This is an example of table footnote.}
\end{minipage}
\end{center}
\end{table}

In general, ML models constitute only a subset of the final system that incorporates the respective ML technology and makes it available to the end users \cite{ml-tech-dept}. With this in mind, the continuous design control approach for handling ML in certified medical systems development fits under the \textit{process management and tools} category. 

Table \ref{table:mitigation} describes how the activities developed as part of the continuous design control help mitigate design problems, such as the accumulation of technical dept, and the role played by the model cards documents as the audit trail of performing these activities.

\subsubsection{Answering regulators' concerns}

Although the use of ML technology within the medical devices is relatively new, the regulators are actively engaged in a dialog with the industry to guide its adoption \cite{GMLP}. This indicates that the regulators are aware of the new technologies, and are considering how to best regulate the development of medical devices that include ML features. Table \ref{table:gmlp} describes how our approach addresses the regulators' concerns.

\begin{table}[p]
\begin{center}
\begin{minipage}{\textwidth}
\caption{Supporting the good ML practice guiding principles \cite{GMLP} with continuous design control}\label{table:gmlp}%
\begin{tabular*}{\textwidth}{p{40mm} p{70mm}}
\toprule
Guiding principles & Implementation \\
\toprule
Multi-disciplinary expertise is leveraged throughout the total product life cycle & The pull request is the venue to perform multi-disciplinary reviews during all development stages\\
\midrule
Good software engineering and security practices are implemented & ML development is integrated into the product and software development leveraging best practices and tools \\
\midrule
Clinical study participants and data sets are representative of the intended patient population & Although clinical studies can be seen as partly outside the scope of product development, the model card document can be used to document data collection protocols and data characteristics that are relevant to the intended patient population. In addition, continuous design control promotes traceability from clinical study data sets to the final model  \\
\midrule
Training data sets are independent of test sets & Following the continuous design practice ensures that the test data set is reviewed, versioned and it is not used during model development \\
\midrule
Selected reference data sets are based upon best available methods & If accepted reference data sets are available, their use in the model development should be promoted and documented in the model card document  \\
\midrule
Model design is tailored to the available data and reflects the intended use of the device & Continuous design control ensures, by including the justification in the model card document, that the data sets are enough to satisfy the intended use\\
\midrule
Focus is placed on the performance of the human-ai team & Testing at different development stages (e.g. training, integration, system) ensures that the different stakeholders' interests are captured\\
\midrule
Testing demonstrates device performance during clinically relevant conditions & Continuous testing in staging environments that mimic the clinically relevant conditions\\
\midrule
Users are provided clear, essential information & The regulatory required documentation contains information that can be collected in the model card documents during development. The continuous design and review activities ensures that the end user documentation fits their needs \\
\midrule
Deployed models are monitored for performance and re-training risks are managed & The monitoring stage of the MLOps pipeline, implementing common techniques for detecting model performance anomalies, together with the regulatory required post-market monitoring practices enable the manufacturer to identify deviations and perform the necessary corrective actions \\
\botrule
\end{tabular*}
% \footnotetext{Source: This is an example of table footnote. This is an example of table footnote.}
% \footnotetext[1]{Example for a first table footnote. This is an example of table footnote.}
% \footnotetext[2]{Example for a second table footnote. This is an example of table footnote.}
\end{minipage}
\end{center}
\end{table}

\subsection{Model card metadata as audit trail}

The emerging model card ecosystem increases the engineering maturity of machine learning model development. The model card metadata document provides an extensible machine-readable medium in which concerns related to the model development can be captured. In our work, we leveraged the model card metadata document to capture the regulatory aspects, such as intended use, the sources of data used for training, or the clinical performance evaluation, relevant when the machine learning model is used in a certified medical device. 

In doing so, we refined existing properties defined in the TensorFlow's model card metadata schema and added new properties when needed (e.g. \texttt{model\_parameters.data[].x\_sources}). We found that the \texttt{description} properties defined in the schema document as string are not structured enough and we used Markdown to have a template driven representation for the intended use of the model (e.g. \texttt{model\_details.documentation}), and for the description of the datasets used in model training (e.g. \texttt{model\_parameters.data[].description}). \new{The approach allowed us to iterate fast, enabling the team members to focus on adding content. The experience, backed with feedback from other implementers, will allow us to identify the relevant information that can eventually be extracted and formalized into model card schema extensions.}

The combination of using a structured document with the semi-structured markdown description is appropriate for the target audience formed by engineers and regulatory professionals, each category contributing using specific modalities. Although the approach is effective at collecting the needed information that serves as an audit trail of the activities performed by the team members, editing the metadata document using text editors does not provide the best user experience for all users.

In the future, we plan to explore with having dedicated editors for regulatory professionals so that they can introduce their content using more familiar approaches, such as \textit{what you see is what you get} editors.

%Link it to the calm compliance paper... 

\subsection{Pull request as continuous design control}

The pull request is the practice typically used by software development teams to manage changes. Our proposal extends the use of pull request throughout the machine learning development lifecycle. Besides software engineers, data engineers and data scientist use the pull request to manage the evolution of the software products within their area of responsibility. As the pull request is linked with requirements (e.g. design inputs), and the introduced changes consist mainly of the machine learning model and the model card metadata (e.g. design outputs), we have an effective quality gate that ensures that design reviews are performed systematically and the appropriate audit trails are build at every iteration throughout the development lifecycle.

\new{\subsection{Handling model anomalies}}

\new{One of the critical advantages of utilizing interpretable machine learning models is that they allow for more efficient anomaly detection and analysis. The ML model can be interpreted as consisting of different components, such as inputs, features, parameters, and weights, and the understandability of the model increases if it can be decomposed into different explainable parts \cite{lipton2018mythos}. Our proposed approach provides a solid foundation to document different model aspects to support explainability, which can, in turn, help the engineering team to perform anomaly and root cause analysis activities. Furthermore, even if initially designed to promote regulatory activities in the form of an audit trail, the fine-grained traceability provides additional support for the anomaly analysis. Based on the results, the team can determine the appropriate corrective actions needed to be performed and included in subsequent model releases. }

\new{\subsection{Safe continuous self-learning}}

\new{The ability to learn after being deployed to real-world use is undoubtedly one of the critical differences between an AI/ML-enabled system and a more traditional rule-based system. However, as discussed previously, due to the current regulatory uncertainties, manufacturers of medical device AI/ML-enabled systems may prefer such AI/ML models that can be deployed in a locked state. It is evident that such a design approach can seriously reduce the benefits of AI/ML-enabled technology. Therefore, alternative yet patient safety ensuring design and development methods are needed.} 

\new{
A robust and effective risk management process is the basis of safe medical device software development. As the process starts with risk identification \cite{iso14971}, the development team must be competent in assessing the product's specific ML change-related aspects, particularly when the chosen technology's complexity and opaqueness increase. In practice, a cross-functional development team should include knowledgeable and experienced data scientists, in addition to the typical set of clinical and product development specialists.
} 

\new{
According to the regulations, medical device manufacturers must seek approval for changes to the approved design of a device prior to making the change, where the change has a substantial impact or can affect the device's conformity with the general safety and performance requirements \cite{mdr}. Therefore, it is clear that if enabled, self-learning can only occur within a pre-determined tolerance and change control plan. In addition, the manufacturer is responsible for demonstrating that the change tolerance complies with the device's intended use, use environment, user groups, and other medical claims prior to placing the device on the market. 
}

\new{
Finally, an essential aspect of self-learning and safety is the ability to monitor the device's performance as a part of the device's post-market surveillance activities. Contrary to the first thought, it can be argued that self-learning AI/ML systems are, in fact, more tolerant against model drift than the locked systems as they are constantly improving their performance with the new data. However, monitoring the constantly changing system can be more difficult as there are additional aspects to consider and measure. The most important thing is to ensure that the device's performance cannot decrease due to an upgrade. 
}

\new{\subsection{Cross domain terminology challenges}}

\new{
A conflicting terminology is a common problem when several domains -- such as medical device regulatory concepts, data science, and software engineering -- are combined within a single project. This problem can lead to miscommunication, misunderstandings, and, at worst, poor decision-making \cite{vogel2011}. The terminology conflicts were also emergent within this paper's context, particularly regarding the term validation. Within the field of ML alone, the term has been used with two different meanings: for data curation (i.e., data validation) or ML model tuning. To make matters even more complicated, in the context of medical device development, validation means confirmation that the particular requirements for specific intended use can be consistently fulfilled \cite{imdrf-ml-terms}. As a practical solution, we propose favoring regulatory terminology in the documents that demonstrate conformity, and, in general enforcing explicit communication to avoid confusion. 
}

\subsection{Information security considerations}

The model card metadata document serves as an effective audit trail of the model development. As such, it contains a plethora of information that should be considered private, as it might contain personal data, information that is not open to public, or even critical trade secrets. While the document should serve as input for generating the public technical documentation of the medical device, as expected by regulation and applicable standards, manufacturers should employ the necessary information management practices to ensure that the properties classified as private are not included in the model card representations intended for public consumption.

\subsection{Limitations}

Our implementation of the proposed approach leverages existing tools and processes widely used by software development teams, such as Git for version control, issues for tracking requirements and work items, or pull requests for reviews and change management. However, we wish to point out that our implementation has not been exposed to a wide range of real life medical products, except Oravizio. For example, managing the evolution of the medical product has been implemented using the feature-branch approach, in which a new branch is created from the mainline, for each requirement, and merged following a successful review. Other development models such as trunk-based development \cite{jorgensen2001putting} have not been investigated thoroughly, although equivalent review facilities are supported by tools used for this development strategy. Therefore, the proposed approach is not intended to be a model that suits any medical product or  situation, which one must follow in a verbatim fashion. Rather, we want to emphasize that systematic reviews and using the model card as the audit trail of regulatory activities represent an effective form of design control that is compatible with the regulatory requirements that govern medical devices that contain software. Similar approaches to track the model card metadata and performing equivalent activities will most likely result in a satisfactory solution from a regulatory perspective.

Operations related to data have been overlooked in the paper, because much of the work happens in data engineers' own environment, following their own ways of working \cite{aho2020demystifying}. However, we proposed model cards as a mechanism to record the trail of provenance from data operations to the model, so that this part can be included in the MLOps pipeline as well. Therefore, exploring the data operations and their relation to model cards is a part of future work we plan to carry out.

\section{Conclusions}

Software engineering industry has widely adopted continuous development and deployment of new features. These features may include AI/ML functions, which have become commonplace in numerous applications, calling for deployment pipelines where such functions can be included in mainstream development activities. Such continuous setup forms a sharp contrast to the development of medical systems, where design controls are often interpreted to require waterfall-like development approach.

In this paper, we propose using an approach where continuous design control for ML is enabled while developing medical systems. The proposed approach builds on our earlier work on MLOps, but extends it with the design controls that are explicitly included in the MLOps pipeline. The approach was demonstrated with an industry system, which is in active use. As future work, we plan to investigate data operations, related to building ML models, in more depth. 

%\backmatter
%
%\bmhead{Supplementary information}
%
%If your article has accompanying supplementary file/s please state so here. 

\bmhead{Acknowledgments}

The authors wish to thank project AHMED and associated consortium, funded by Business Finland, for supporting this research.

\bmhead{Data Availability Statement}

Data sharing not applicable to this article as no datasets were generated or analysed during the current study.

\bmhead{Conflict of interest}
The authors declare that they have no conflict of interest.

\end{document}